\documentclass[aps,prb,twocolumn,notitlepage,superscriptaddress,showpacs]{revtex4-1}

\usepackage{amsmath,latexsym,amssymb,bm}
\usepackage{hyperref}
\usepackage{color}
\usepackage{graphicx,subfigure}
\usepackage{multirow}

\begin{document}
\title{Pairing versus phase coherence of doped holes in distinct quantum spin backgrounds}

\author{Zheng Zhu}
\affiliation{Department of Physics, Massachusetts Institute of Technology, Cambridge, MA, 02139, USA}
\author{D. N. Sheng}
\affiliation{Department of Physics and Astronomy, California State University, Northridge, CA, 91330, USA}
\author{Zheng-Yu Weng}
\affiliation{Institute for Advanced Study, Tsinghua University, Beijing, 100084, China}
\affiliation{Collaborative Innovation Center of Quantum Matter, Tsinghua University, Beijing, 100084, China}

\date{\today}

\begin{abstract}
We examine the pairing structure of holes injected into two \emph{distinct} spin backgrounds: a short-range antiferromagnetic phase versus a symmetry protected topological phase. Based on density matrix renormalization group (DMRG) simulation,  we find that although there is a strong binding between two holes in both phases,  \emph{phase fluctuations} can significantly influence the pair-pair correlation depending on the spin-spin correlation in the background. Here the phase fluctuation is identified as an intrinsic string operator nonlocally controlled by the spins. We show that while the pairing amplitude is generally large, the coherent Cooper pairing can be substantially weakened by the phase fluctuation in the symmetry-protected topological phase, in contrast to the short-range antiferromagnetic phase. It provides an example of a non-BCS mechanism for pairing, in which the paring phase coherence is determined by the underlying spin state self-consistently, bearing an interesting resemblance to the pseudogap physics in the cuprate.

\end{abstract}

\maketitle

\emph{Introduction.---}In the conventional  Bardeen-Cooper-Schrieffer (BCS) theory of superconductivity\cite{BCS}, quasiparticles start to form Cooper pairs at the transition temperature $T_c$, where they simultaneously achieve long-range phase coherence. In strongly correlated systems, the size of a Cooper pair is expected to be much reduced, which may be meaningfully isolated from each other and even survive above $T_c$ without phase coherence~\cite{Kivelson1995}. The low-temperature pseudogap regime of the high-$T_c$ cuprate superconductor is widely believed to be the case~\cite{Kivelson1995,Lee2006}. Nevertheless, the pairing mechanism as well as the relationship between pseudogap and superconductivity are still unsettled \cite{Lee2006, Anderson1987,Anderson,Yazdani2008,Shen2014,Keimer2015,Fabio2016, Plakida2006, Demler1998, Paramekanti2004, Anderson2006,Weng2011,Scalapino2012, Gull2014,Sakai2016,Loret2016,Sorella2002,Dolfi2015,Lee2014,White1997}.

The standard understanding of the non-phonon pairing in a doped Mott insulator/antiferromagnet is based on the resonating-valence-bond (RVB) picture~\cite{Anderson1987,Anderson}, where the spin background is composed of condensed spin RVB pairs such that doped charges have an energetic incentive to pair. Identifying the pairing of two holes in the background of correlated quantum spins is therefore an important step to investigate  superconductivity in a strongly correlated system. In particular, it would be highly instructive to artificially change the nature of the spin background in order to fully expose the underlying complexity of the pairing mechanism in a doped Mott insulator.  After all, the pseudogap phase \cite{Keimer2015} in the cuprates arises above $T_c$ where the spin background as thermally excited states is already quite different from the ground state. Experimentally, different spin backgrounds may be also realizable in cold atom systems by loading quantum gas into optical lattices \cite{Esslinger2010, Hilker2017,Zwierlein2015,Greiner2016,Bakr2016,Boll2016,Sherson2010}.
\begin{figure}[bh]
\begin{center}
\includegraphics[width=0.40\textwidth]{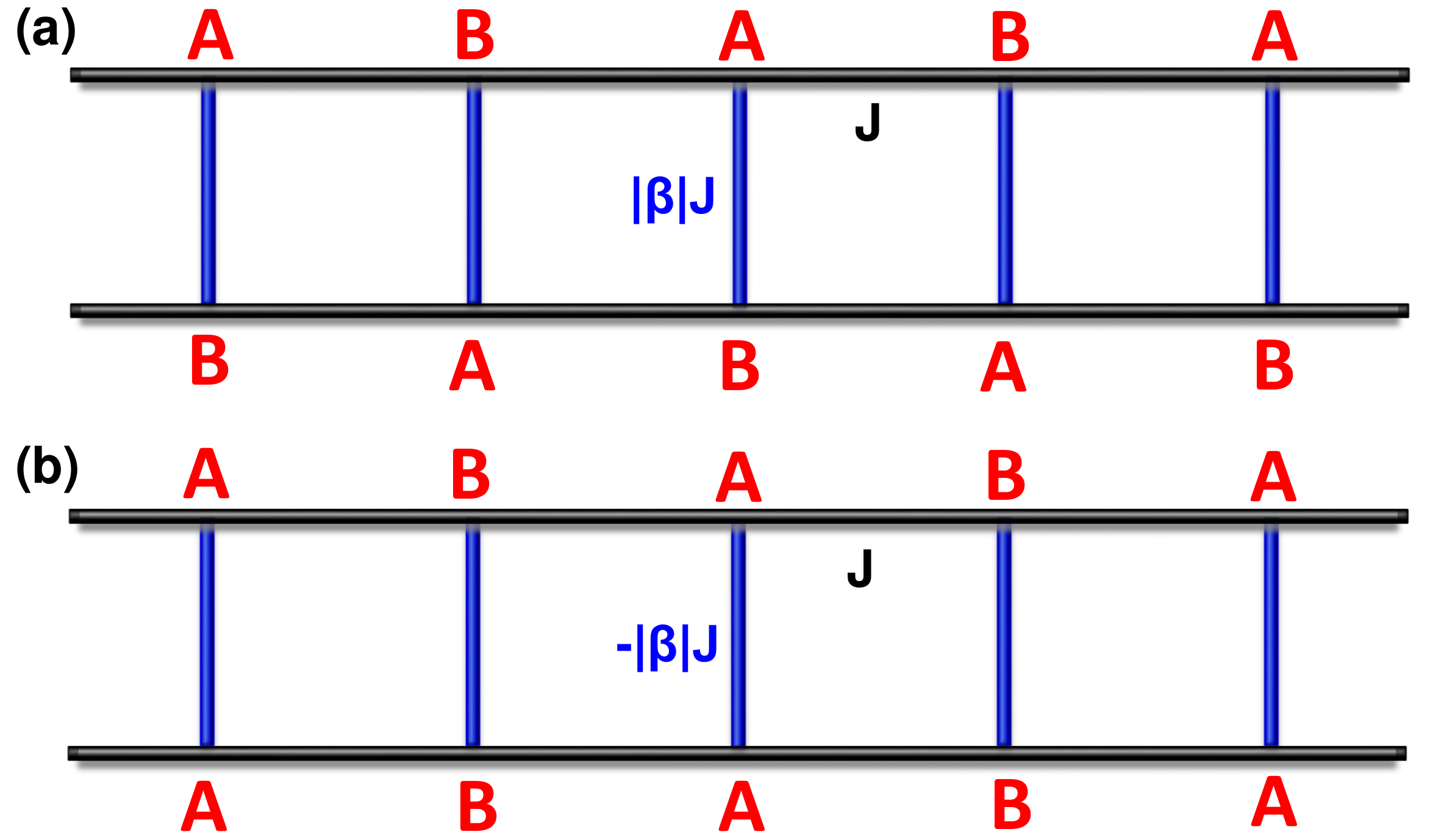}
\end{center}
\par
\renewcommand{\figurename}{Fig.}
\caption{(Color online) Distinct quantum spin backgrounds can be realized in a two-leg spin-1/2 ladder with the superexchange couplings along the leg, $J$, and the rung, $\beta J$: (a) $\beta>0$; (b) $\beta<0$ (see the text).  Two types of sublattices, A and B, can be defined to incorporate the Marshall sign structure~\cite{Marshall1955} in the corresponding ground state [cf. Eq. (\ref{phi0}) in Appendix A]. }
\label{sFig1}
\end{figure}

\begin{figure*}[th]
\begin{center}
\includegraphics[width=1.0\textwidth]{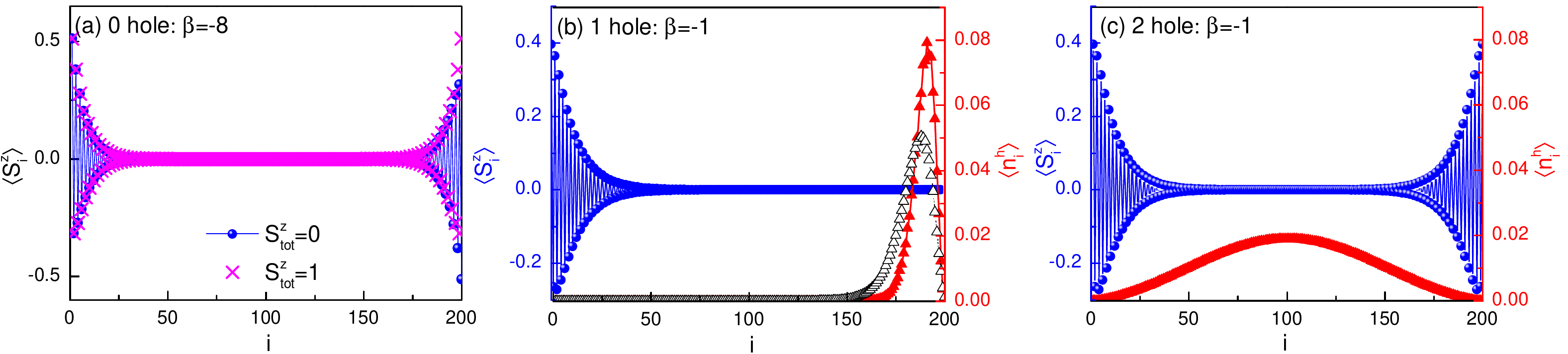}
\end{center}
\par
\renewcommand{\figurename}{Fig.}
\caption{(Color online) (a) The distribution of spin $\langle S^z_i\rangle $ at half-filling ($\beta=-8$) obtained by DMRG. There are four-fold degeneracies with an $S=1/2$ edge spin at each side of the open boundary (blue circle and red cross); (b) One injected hole at $\beta=-1$ is trapped at the boundary (right-hand-side) as indicated by $\langle n^h_i\rangle $ (red triangle: $H$; open triangle: $H_0$); (c) Two injected holes form a bound pair, which becomes mobile in the bulk with an extended density profile (red triangle, indistinguishable between $H_0$ and $H$), while two $S=1/2$ spin edge modes reemerge at the boundaries.
}
\label{Fig1}
\end{figure*}

In this paper, we study the ground state of two holes injected into a spin background which can be continuously tuned between two distinct phases: a short-range antiferromagnet~\cite{Dagotto1996} phase and a topological Haldane/Affleck-Kennedy-Lieb-Tasaki (AKLT) phase~\cite{Haldane1983, Affleck1987} with edge modes. By using density matrix renormalization group (DMRG) and analytic methods, we show that, once the spin-spin correlations become short ranged, a strong pairing of two holes does take place in both systems. We find that the binding energy is composed of two parts, with the kinetic energy term making an additional significant contribution besides the well-known RVB mechanism related to the pairing due to the superexchange coupling. However, this strongly binding  of two holes is shown to be only responsible for a large \emph{pairing amplitude} in the ground state. A Cooper pairing further involves a phase component, whose generic form is identified analytically, and it can effectively distinguish the two underlying spin backgrounds. The significant fluctuations of the phase component strongly suppress the pair-pair correlations of the Cooper pairs particularly in the AKLT state. It implies that upon introducing a finite density of holes, the ground state may not be BCS-like by nature, but rather with a preformed/composite pairing structure which allows for phase-disordering controlled by the underlying spin correlations in a doped Mott insulator. Its implications for the pseudogap phase will be discussed.

\emph{The model.---}Distinct quantum spin backgrounds can be realized in an anisotropic Heisenberg spin-1/2 Hamiltonian on a two-leg square lattice ladder  [see Fig. ~\ref{sFig1}],
\begin{equation} \label{eq:HJ}
 H^{\beta}_J = J \sum_{i} \left(  {\mathbf S}_{1i}\cdot{\mathbf S}_{1i+1}+ {\mathbf S}_{2i}\cdot{\mathbf S}_{2i+1}\right) +\beta J \sum_{i} {\mathbf S}_{1i}\cdot{\mathbf S}_{2i},
\end{equation}
where the subscripts, $1$ and $2$, label the leg numbers. Note that the anisotropic parameter $\beta$ denotes the ratio of interchain and intrachain coupling with $-\infty< \beta<+\infty$, and it changes sign at $\beta=0$ where the two chains are decoupled, which separates two distinct spin phases as to be discussed below. In particular, two types of sublattices, A and B, can be defined at $\beta>$ and $<0$,  as shown in Fig. ~\ref{sFig1}, to incorporate the Marshall sign structure~\cite{Marshall1955} in the corresponding ground states [cf. Eq. (\ref{phi0}) in Appendix A]. In this paper, we set $J=1$ as the unit of energy.

Then, holes are injected into the half-filled spin background. For simplicity of analytic analysis, we consider the hopping of doped holes \emph{only} along the chain direction with the interchain hopping $t_{\perp}= 0$, i.e.,
\begin{align}\label{Htj}
H_{t_{\parallel}} &= - t\sum_{i\sigma}\left( c_{1i\sigma}^\dagger c_{1i+1\sigma}+ \ c_{2i\sigma}^\dagger c_{2i+1\sigma} + \mathrm{h.c.}\right)~.
 \end{align}
Here, the no-double-occupancy constraint, i.e., $\sum_{\sigma}c_{1,2 i\sigma}^\dagger c_{1,2 i\sigma} \leq 1$, is enforced in the Hilbert space. The resulting Hamiltonian $H$ is given by
\begin{equation}\label{H}
 {H}\equiv {H}_{t_{\parallel}} + {H}^{\beta}_J~.
 \end{equation}

In order to understand the pairing mechanism properly in the above $t$-$J$ type model, we further make a comparison study by modifying the kinetic energy (hopping) term with inserting a spin dependent sign $\sigma=\pm 1$:
\begin{align}\label{Hst}
H_{\sigma \cdot t_{\parallel}} &= - t\sum_{i\sigma} \sigma\left(  c_{1i\sigma}^\dagger c_{1i+1\sigma}+ \ c_{2i\sigma}^\dagger c_{2i+1\sigma} + \mathrm{h.c.}\right)~.
 \end{align}
 The resulting Hamiltonian $H_0$ is given by
\begin{equation}\label{H0}
 {H}_0\equiv {H}_{\sigma \cdot t_{\parallel}} + {H}^{\beta}_J ~,
 \end{equation}
known as the $\sigma$$\cdot$$t$-$J$ model \cite{ZZ2013}, in which the introduced ``spin-orbit coupling'' or sign $\sigma$ can effectively remove an intrinsic frustration hidden in the $t$-$J$ model, subsequently affecting the pairing structure drastically. Physically, such intrinsic frustration in the $t$-$J$ type hopping term corresponds to a sequence of signs, known as the phase string \cite{Sheng1996,Weng1997,Wu2008,Zaanen2011}, picked up by the hole as the result of disordered Marshall signs in the spin background  (cf. Appendix A).

\emph{DMRG results.---}At $\beta>0$, $ H^{\beta}_J $ hosts a short-range (spin gapped) antiferromagnetic ground state, while at $\beta<0$, the ground state becomes topological with a gapped bulk state and  gapless edge modes.   One can find a four-fold degeneracy with two $S=1/2$ edge spins trapped at two ends of boundaries, leading to the degenerate states with total spin $S_{\emph {tot}}^z=0, 1$ under an open boundary condition (OBC), which is the same as expected in an $S=1$ Haldane phase~\cite{Haldane1983} or AKLT-type ground state~\cite{Affleck1987} [cf. Fig.~\ref{Fig1}(a)].
Here $\beta=0$ is indeed a quantum critical point~\cite{Senthil2004}, where the system reduces to two decoupled chains of gapless spin Luttinger liquids~\cite{Tomonaga1950,Luttinger1963} at half-filling.

\begin{figure}[bhp]
\begin{center}
\includegraphics[width=0.4\textwidth]{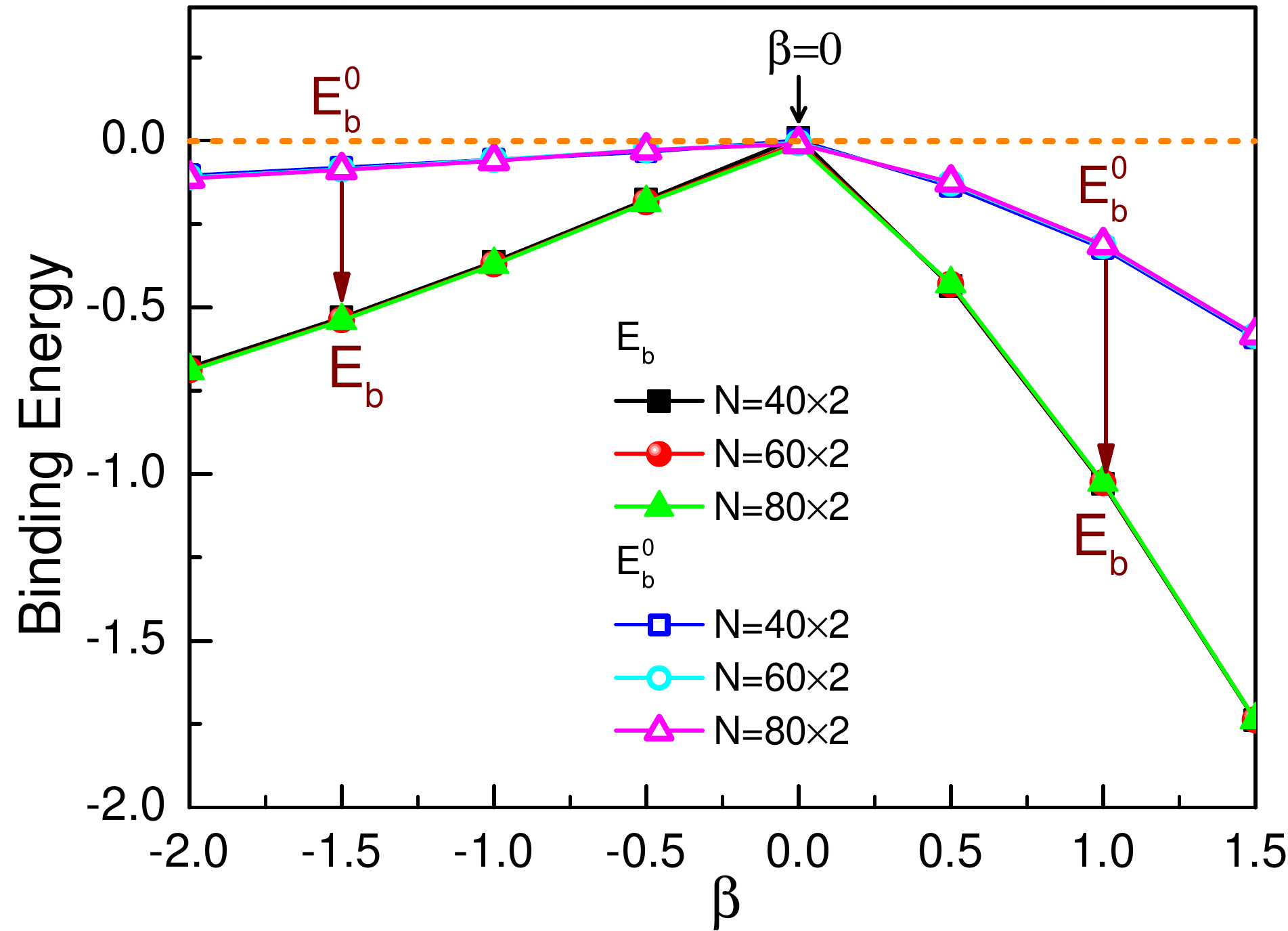}
\end{center}
\par
\renewcommand{\figurename}{Fig.}
\caption{(Color online). Two-hole binding energies, $E_b^0$ and $E_b$, for $H_0$ and $H$, respectively, as a function of $\beta$. The significant enhancement of the binding strength for $H$ is indicated.}
\label{Fig2}
\end{figure}


Figure~\ref{Fig1}(b) illustrates that a single hole injected into the half-filling Haldane phase at $\beta<0$ is trapped at the boundary for both $H$ and $H_0$. Note that the $S^z=\pm 1/2$ profile disappears on the side where the hole is trapped, suggesting that the spin edge mode is bound to the doped hole to form an emergent spinless charge mode (holon) at the boundary. However, once two holes are injected into the system, they further form a \emph{mobile} bound pair in the bulk as illustrated in Fig. \ref{Fig1}(c) rather than trapping at boundaries, indicating that binding forces between two holes emerge for both $H$ and $H_0$.

In the following, we focus on the case of two doped holes. In general, one may define the two-hole binding energy $E_b$ as $E_b=E^{2h}_G-E_G^{0}-2(E^{1h}_G-E_G^{0})$,
where $E^{2h}_G$,  $E^{1h}_G$, and  $E^{0}_G$ denote the ground state energies of two-hole, one-hole, and half-filling states, respectively. As shown in Fig. ~\ref{Fig2}, the two holes indeed form a bound state at either $\beta>0$ or $\beta<0$. However, as shown in Fig. ~\ref{Fig2}, the total $E_b$ for $H$ is \emph{substantially lower} than the binding energy $E^0_b$ for $H_0$ on both sides of $\beta\neq 0$. This means that the kinetic energy must play a crucial role in the binding energy,  because the superexchange term $ {H}^{\beta}_J$ are the same for both $H$ and $H_0$.

\begin{figure}[tbp]
\begin{center}
\includegraphics[width=0.43\textwidth]{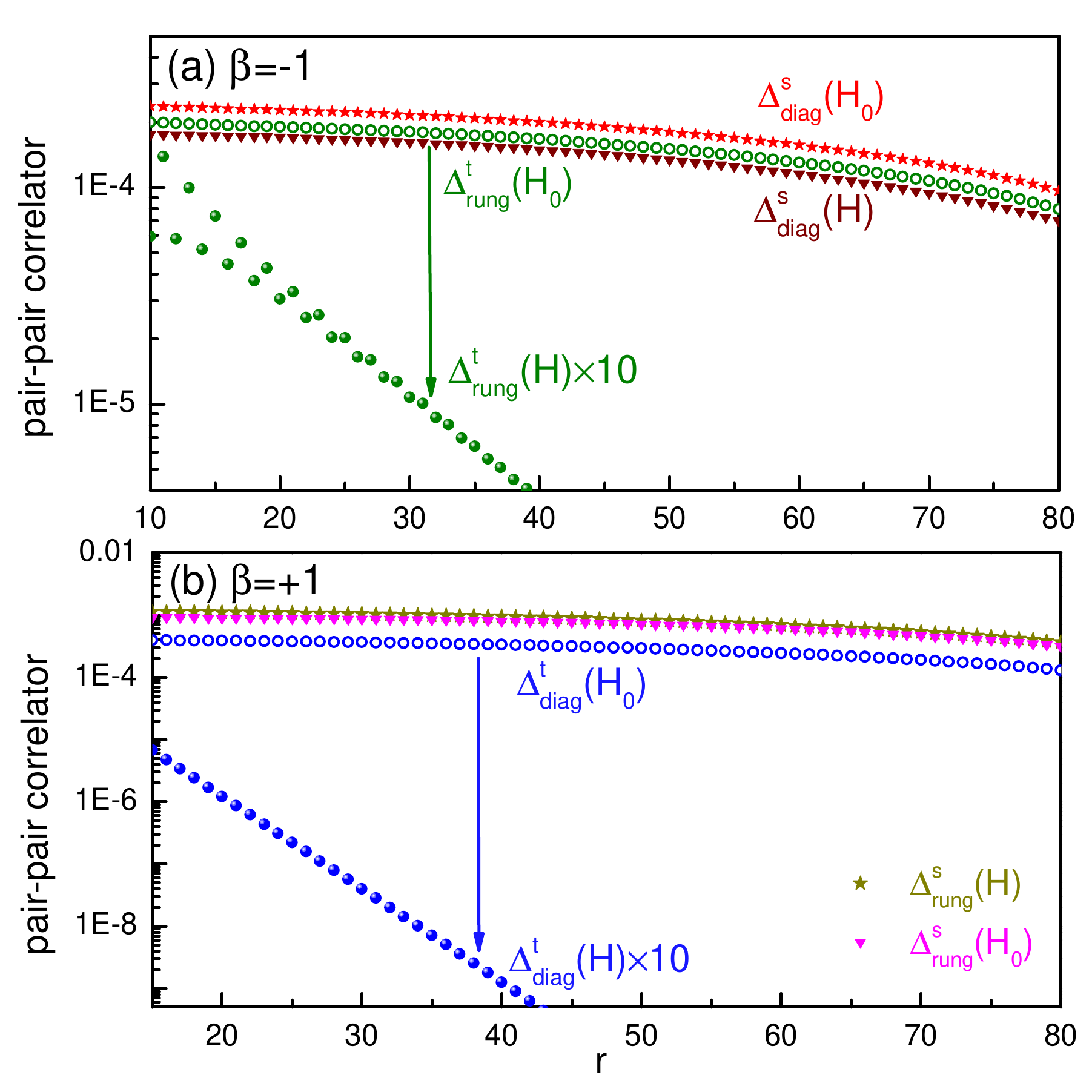}
\end{center}
\par
\renewcommand{\figurename}{Fig.}
\caption{(Color online)  The Cooper pair-pair correlators for $\beta=-1$ (a) and $\beta=+1$(b), governed by $H_0$ and $H$ in singlet and triplet channels with the system size $N=200\times2$. ${\Delta}^{s,t}_{\mathrm{rung}}$ and ${\Delta}^{s,t}_{\mathrm{diag}}$ denote Cooper pairs on the rung and diagonal bonds (see text), respectively. }
\label{Fig3}
\end{figure}

Alternatively, one may further examine the pair-pair correlators, $\left\langle \hat{\Delta}^{s,t}_{ij} \left(\hat{\Delta}^{s,t}_{i'j'}\right)^{\dagger}\right\rangle $, as shown in Fig.~\ref{Fig3}, in which the Cooper pair operators in the singlet/triplet channels are defined as follows:
\begin{align}\label{spair}
\hat{\Delta}^s_{ij} &=\frac 1 {\sqrt{2}}\sum_{\sigma}\sigma {c}_ {1 i \sigma} { c}_ {2 j -\sigma}~,  \\
\hat{\Delta}^t_{ij} & \equiv \frac 1 {\sqrt{2}}\sum_{\sigma} {c}_ {1 i \sigma} { c}_ {2 j -\sigma} \label {tpair}~.
\end{align}
Due to the short-range nature of the pairing, here we focus on the local rung ($j=i$) pairing (denoted by ${\Delta}^{s,t}_{\mathrm{rung}}$ in Fig.~\ref{Fig3}) and the next nearest neighbor ($j=i\pm 1$) diagonal pairing (${\Delta}^{s,t}_{\mathrm{diag}}$) at two opposite legs. However, as indicated in Fig.~\ref{Fig3}, the overall pair-pair correlations are barely enhanced in the ground state of $H$ as compared to that of $H_0$.
In fact, at $\beta=-1$, one finds a slight reduction in the singlet diagonal pairing but a substantial decrease to an exponential decay in the triplet rung channel [indicated by the arrows in Fig.~\ref{Fig3} (a)]. At $\beta=1$, the singlet rung pairing are almost the same, whereas the triplet diagonal pairing strongly reduces to an exponential decay [cf. Fig.~\ref{Fig3} (b)] for the $H$ model.

This seems contradictory sharply to the substantial enhancement in the binding energy of $H$ shown in Fig.~\ref{Fig2}. The opposite trends in the binding energy vs. the pair-pair correlation in the above measurements clearly indicate that the ground state of $H$  must be fundamentally different from simply creating a coherent Cooper pair of holes on a half-filled spin background as in $H_0$. The latter case is discussed in Appendix A2, where it is shown that the Cooper pairing is coherent in the ground state of $H_0$ and the hole pairing can thus serve as a good reference in Figs.~\ref{Fig2} and \ref{Fig3} to understand the nature of pairing described by $H$. 
\begin{figure}[tbp]
\begin{center}
\includegraphics[width=0.43\textwidth]{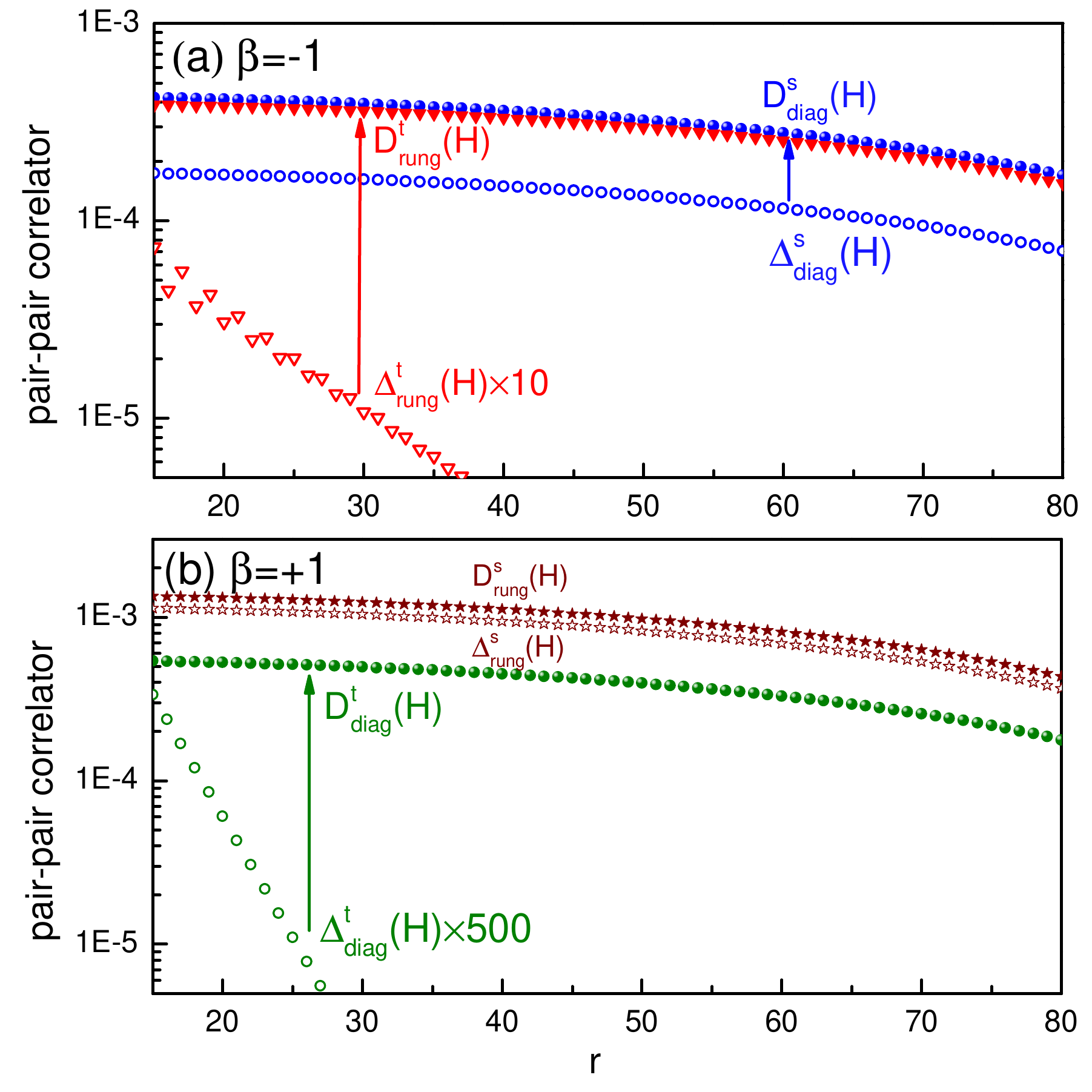}
\end{center}
\par
\renewcommand{\figurename}{Fig.}
\caption{(Color online)  The much enhanced correlators of new pairing operators $\hat{D}^{s,t}_{ij} $ defined in Eq. (\ref{cc}) vs. the Cooper pairs [Eqs. (\ref{spair}) and (\ref{tpair})], which are governed by $H$ at $\beta=-1$ (a) and $\beta=+1$(b) with the system size $N=200\times2$. }
\label{Fig4}
\end{figure}

\emph{Microscopic origin of the enhanced binding in $H$.---}To compare $H$ with $H_0$, one may introduce a unitary transformation
\begin{equation}\label{unitary}
e^{i\hat{\Theta}}\equiv e^{-i\sum_{\alpha i}n_{\alpha i}^h\hat{\Omega}_{\alpha i}}
\end{equation}
to formally transform ${H}_{t_{\parallel}} $ [Eq.~(\ref{Htj})] into the ``sign-free'' $H_{\sigma\cdot t_{\parallel}} $ [Eq. (\ref{Hst})]. Here $n_{\alpha i}^h$ denotes the hole number at leg $\alpha=1$, $2$ and site $i$ along the chain.  On the other hand, under such a transformation, ${H}^{\beta}_J $ turns into ${H}^{\beta}_J+{H}_I $ with an extra term
\begin{equation}\label{HI}
 {H}_I = -\frac J 2 \beta \sum_{i}\left\{ \left[1-e^{-i\pi \sum_{l<i}(n^h_{1l}-n^h_{2l})} \right] { S}^+_{1i}{S}^-_{2i}+h.c.\right\}~.
 \end{equation}
Consequently one finds
\begin{equation}\label{tildeH}
e^{-i\hat{\Theta}}  H  e^{i\hat{\Theta}}  =  {H}_0 +{H}_I~.
 \end{equation}
Therefore, the difference between $H$ and $H_0$ is now represented by ${H}_I$ in the new Hamiltonian (\ref{tildeH}) which is due to the nontrivial sign structure (phase string \cite{Sheng1996,Weng1997,Wu2008,Zaanen2011}) hidden in $H$ that cannot be truly ``gauged away'' by the unitary transformation (\ref{unitary}).

It can be shown that ${H}_I$ gives rise to a linear-confining potential if a single hole behaves coherently as a Bloch wave (cf. Appendix B). However, once two holes are injected into the spin ladder, the above confinement potential disappear if the two holes form a tightly bound pair. Indeed, if the two holes sit at the same rung $l$ of different chains, e.g., $n^h_{1l}=n^h_{2l}=1$, the potential vanishes in ${H}_I$.
If the two holes are separated along the chain direction by a distance $l_{ij}$, then a pairing potential arises, which is linearly proportional to $J|\beta|l_{ij}$ (cf. Appendix B3), thus the binding strength of two holes is substantially enhanced in $H$ as compared to $H_0$, as shown in Fig. ~\ref{Fig2}.

Furthermore, similar to the sign-free $H_0$, the pairing as governed by $H_0+H_I$ should also behave \emph{coherently} due to the pair translational symmetry in $H_I$. In other words, a \emph{Cooper pair} should be still coherent and thus constitute a true hole pair creation operator in the transformed \emph{new} ground state  $ |{\Phi}_G\rangle\equiv e^{-i\hat{\Theta}}|\Psi_G\rangle $. Such a pair operator can be denoted by $\hat{D}^{s,t}_{ij} $.

The pair-pair correlations of $\hat{D}^{s,t}_{ij} $ as compared to the corresponding ones of the Cooper pair operators are shown in Fig.~\ref{Fig4}, which are calculated by DMRG (see below). As illustrated in Fig.~\ref{Fig4} , they are indeed enhanced (indicated by the arrows) as compared to those of Cooper pairing, $\hat{\Delta}^{s,t}_{ij} $ shown in Fig.~\ref{Fig3} for both $H$ \emph{and} $H_0$.

\emph{Novel pairing structure in $H$.---}However, in the original $ |\Psi_G\rangle $, such a new coherent pair operator is transformed back to the original pair operator $\hat{\Delta}^{s,t}_{ij} $ by the following generic form:
\begin{align}  \label{cc}
\hat{\Delta}^{s,t}_{ij} =\left(\hat{D}^{s,t}_{ij}  \right) e^{i\Phi^s_{ij}} ~.
 \end{align}
It means that the true pair operator $\hat{\Delta}^{s,t}_{ij} $ has a composite structure obtained by multiplying a pair amplitude $\hat{D}^{s,t}_{ij} $ by a phase factor with $\Phi^s_{ij}\equiv \hat{\Omega}_{1 i}+\hat{\Omega}_{2 j} $ that is controlled by the spin background. Generally speaking, the pair-pair correlation as an equal-time propagator of a Cooper pair not only reflects the binding strength of the holes, but also encodes the phase information of the pair. In the following, we shall see that the phase fluctuations in Eq. (\ref{cc}) becomes a crucial part of the pairing structure.

\begin{figure}[tbp]
\begin{center}
\includegraphics[width=0.43\textwidth]{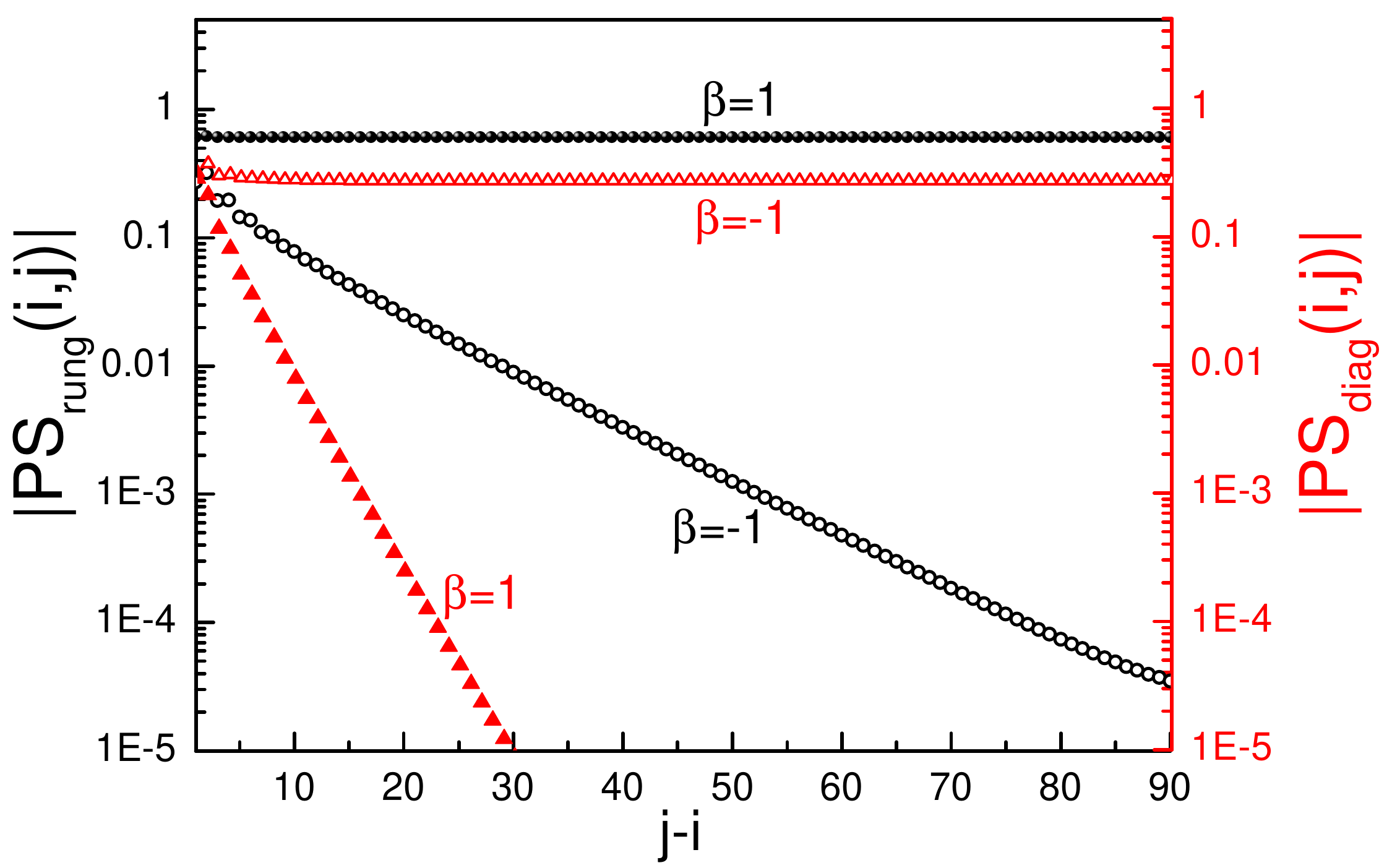}
\end{center}
\par
\renewcommand{\figurename}{Fig.}
\caption{(Color online)The correlations of the rung and diagonal phase-string order parameters in Eq. (\ref{cc}) (see text) clearly distinguish the distinct nature of the underlying spin background ($N=200\times2$).}
\label{Fig5}
\end{figure}

Here the phase operator appearing in $ e^{i\Phi^s_{ij}}$ is given by
 $\hat{\Omega}_{\alpha i}\equiv \pi \sum_{l>i}n_{\alpha l} ^{ \downarrow}$ (see below),
which involves all $\downarrow$-spins (whose number operator is denoted by $n^{\downarrow}_{\alpha l}$ at leg $\alpha=1$, $2$ and site $l$ along the chain) at $l>i$ under the OBC.
For example, at $i=j$, one has
\begin{align}\label{pairph}
e^{i\Phi^s_{ii}} =e^{-i\pi \sum_{l>i}\left( S^z_{1 l}+S^z_{2 l}\right)}~,
\end{align}
which, acting on the half filling background, has precisely the same form as the phase factor in the string operator of the AKLT spin chain \cite{Nijs1989}. We define $\mathrm{PS}_{{\rm{rung}}}({i,j}) \equiv  \left\langle {e^{i\Phi^s_{ii}}e^{-i\Phi^s_{jj}} }\right\rangle$  and  $\mathrm{PS}_{{\rm{diag}}}({i,j})  \equiv  \left\langle {e^{i\Phi^s_{ii+1}}e^{-i\Phi^s_{jj+1}} } \right\rangle$, and they give rise to an exponential decay at $\beta < 0$ and $\beta> 0$, respectively, however, $\mathrm{PS}_{\text{rung}} $ ($\mathrm{PS}_{\text{diag}}$) remains a constant at $\beta>0$ ($\beta <0$), as calculated by DMRG and illustrated in Fig. \ref{Fig5}.
These confirm that the phase component in Eq. (\ref{cc}) does fluctuate strongly and depend sensitively on the spin correlation of the spin background, leading to profound effects on the divergent behaviors in the pair-pair correlations shown in Figs.~\ref{Fig3} and \ref{Fig4}.

Then, what Fig.~\ref{Fig4}(a) clearly demonstrates is that the strongly enhanced binding of holes in $H$ at $\beta=-1$ has been mostly translated into the \emph{pairing amplitudes}, $\hat{D}^{s}_{\mathrm{diag}} $ and $\hat{D}^{t}_{\mathrm{rung}} $, whereas the corresponding Cooper pairings, $\hat{\Delta}^{s}_{\mathrm{diag}} $ and $\hat{\Delta}^{t}_{\mathrm{rung}} $, are substantially reduced by contrast, due to the phase factor shown in Eq. (\ref{cc}).
Energetically, by considering the spin background, the local triplet rung and singlet diagonal bindings are indeed favored at $\beta<0$, while the triplet diagonal and singlet rung bindings \cite{note1} are naturally expected when $\beta>0$.

On the other hand, at $\beta=+1$, the enhancement of $D^{s}_{\mathrm{rung}}(H) $ is only about $18\%$ larger than ${\Delta}^{s}_{\mathrm{rung}} (H) $ as shown in Fig.~\ref{Fig4}(b) and the overall phase average of $\mathrm{PS}_{\text{rung}}$ is finite $\sim 0.6$ (cf. Fig. \ref{Fig5}). It means that the phase fluctuations can get much reduced in a short-range antiferromagnetic background as compared to the AKLT-type of spin state at $\beta<0$, where the phase frustration is much enhanced. The important implication is that the short-range antiferromagnetic spin correlation is actually in favor of the phase coherence of the Cooper pairing in a doped Mott insulator. It is also consistent with the coherent pairing previously studied with $t_{\perp}\neq 0$ by probing the response to an inserting flux in a ring geometry of the two-leg ladder \cite{ZZ2014}.  Therefore, it provides an important hint to the superconducting transition as a phase coherence transition at finite doping. Namely, with the increase of temperature, the short-range antiferromagnetic background may be changed by thermal spin excitations such that the cancellation in the phase of Eq. (\ref{cc}) to realize phase coherence becomes weakened. In other words, with the thermally excited spins, the phase fluctuations could become enhanced to eventually destroy the Cooper pairing and result in the superconducting transition. Although this is beyond the present scope,  our results provide  a toy-model-study by turning $\beta<0$ to show how the pairing phase fluctuation can indeed be increased by driving the spin background away from the short-range antiferromagnetism.

\emph{Conclusion.---}The above combined DMRG and analytic analysis clearly demonstrate that a general non-Cooper-like pairing structure can be obtained in doped Mott insulators in a unified form as given in Eq. (\ref{cc}).
Here, a Cooper pair exhibits an intrinsic phase fluctuation identified by a nonlocal string operator, which sensitively depends on the spin correlation varying from a Haldane-AKLT phase to a short-range antiferromagnet. In the former phase, the pairing amplitude is much stronger whereas the Cooper pairing is significantly reduced due to the phase fluctuation. In contrast, the Cooper pairing is strengthened in the latter phase due to the suppression of the pairing phase fluctuation with the short-range antiferromagnetic spin correlations. This result indicates that a Cooper pair of two holes can be subject to strong phase fluctuations depending on the nature of spin-spin correlation in the pseudogap-like (spin-gapped) background. Previously the coherent part of the pairing has been also studied \cite{ZZ2014} at $\beta>0$ with a finite interchain hopping $t_{\perp}$ in Eq.~(\ref{Htj}). However, in view of the present study, the possible presence of a non-Cooper-pair amplitude needs a further clarification \cite{note2}. In completing this work, a finite doping at $\beta<0$ case was reported recently \cite{Jiang2017}, where a quasi-long-range pair-pair correlation is also found for the singlet Cooper pairing on diagonal bonds, similar to the corresponding channel shown in Fig. \ref{Fig4}. But the singlet Cooper pairing may be \emph{much weakened} as compared to the preformed pair amplitude $\hat{D}$ shown in Fig. \ref{Fig4}, and thus a more careful study of the hidden non-Cooper-pairing amplitude at finite doping would be highly desirable.

\emph{Acknowledgements.---}Useful discussions with Dung-Hai Lee, Hongchen Jiang, and Shuai Chen are acknowledged. This work is supported by Natural Science Foundation of China (Grant No. 11534007), MOST of China (Grant Nos. 2015CB921000 and 2017YFA0302902). D.N. Sheng is  supported by US National Science Foundation Grant  DMR-1408560.

\appendix

\section{Precise sign structure of $H_0$ and $H$}
\label{sec:}

\subsection{Sign structure of $H^{\beta}_J$: Two distinct quantum spin phases at half-filling}

Two distinct spin phases at half-filling can be separated by the quantum critical point at $\beta=0$ as governed by $H^{\beta}_J$ in Eq.(\ref{eq:HJ}), in which the anisotropic parameter $-\infty< \beta<\infty$, changing sign at $\beta=0$ where the two chains are decoupled.  In the following, we show that they can be distinguished by two distinct sign structures.

At $\beta\ge 0$, the ground state wavefunction of $H^{\beta}_J$ [Eq.(\ref{eq:HJ})] on a bipartite lattice satisfies the so-called Marshall sign structure~\cite{Marshall1955}. Namely, the half-filling ground state $|\Psi_G\rangle_{\emph{half-filling}}$ can be expressed by
\begin{align} \label{phi0}
|\Psi_G\rangle_{\emph{half-filling}}&=  (-1)^{\hat{N}^{\downarrow}_{A}}|{\Phi}_G\rangle_0, \nonumber \\
|{\Phi}_G\rangle_{0}&=\sum_{\chi}c_{\chi}|\chi\rangle,
\end{align}
where $c_{\chi}\geq 0$ and $|\chi\rangle $ denotes an Ising basis of spins. Here, the sign factor $(-1)^{\hat{N}^{\downarrow}_{A}}$ is known as the Marshall sign structure ~\cite{Marshall1955} in which $\hat{N}^{\downarrow}_{A}$ measures the total number of $\downarrow$-spins at the $A$-sublattice indicated in Fig.~\ref{sFig1}(a).

Now we generalize the Marshall sign structure to the case of $\beta<0$ by redefining two sub-lattices in Fig.~\ref{sFig1}(b), which is different from the case for $\beta>0$ in Fig.~\ref{sFig1}(a). Then by introducing the following unitary transformation: $ \tilde{H}^{\beta}_J \equiv (-1)^{\hat{N}^{\downarrow}_{A}}{H}_J (-1)^{\hat{N}^{\downarrow}_{A}}
$, one finds
\begin{align} \label{tildeHJ}
 \tilde{H}^{\beta}_J
 &=J \sum_{i} \left(  {S}^z_{1i}{ S}^z_{1i+1}+ {S}^z_{2i}{ S}^z_{2i+1}\right) +\beta J \sum_{i} { S}^z_{1i}{S}^z_{2i}\nonumber \\
 &-\frac J 2 \sum_{i} \left(  {S}^+_{1i}{ S}^-_{1i+1}+ {S}^+_{2i}{ S}^-_{2i+1}+|\beta|  { S}^+_{1i}{S}^-_{2i}+h.c. \right).
\end{align}
In the Ising basis $\{|\chi\rangle \}$, all the off-diagonal matrices of  $ \tilde{H}^{\beta}_J$ are therefore negative-definite:   $\langle \chi'|\tilde{H}^{\beta}_J|\chi\rangle \leq 0 $. According to the Perron-Frobenius theorem, one has the general form of the ground state (\ref{phi0}). Therefore, the ground states both satisfy the Marshall sign rule in Eq. (\ref{phi0}) with different definitions of the sublattices at $\beta> 0$ and $\beta<0$, respectively.

Figure 2 (a) in the main text shows the spin $\langle S^z_i\rangle $ distribution of the ground states at $\beta=-8$ obtained by the density matrix renormalization group (DMRG) method under open boundary condition (OBC). One can find there are four-fold degeneracies with two $S=1/2$ edge spins trapped at two ends of boundaries, leading to the degenerate states with total spin $S^{\emph {tot}}_z=0, 1$ under OBC as expected in an $S=1$ AKLT ground state.

\subsection{Sign free $H_0$ and coherent Cooper pairing upon doping}

In the above, we have seen that $H^{\beta}_J$ is a \emph{sign free} Hamiltonian [cf. Eq.(\ref{tildeHJ})] and there is no nontrivial sign structure in the ground state (\ref{phi0}) except the Marshall sign rule~\cite{Marshall1955}.

Now we consider injecting holes into the half-filled spin backgrounds. First, the Ising spin basis $|\chi\rangle $ in Eq. (\ref{phi0}) can be changed to the Ising-spin-hole basis $|\chi; \{l_h\}\rangle $:
\begin{equation}\label{basis2}
|\chi; \{l_h\}\rangle \equiv c_{l_1\sigma_1}c_{l_2\sigma_2}...|\chi\rangle
\end{equation}
where  $\{l_h\}=l_1<l_2<....$ denote the sequence of an arbitrary hole configuration (for simplicity, here we omit the leg index since there is no interchain hopping), and $\chi$ a configuration of the Ising spins. Here we always consider the two-leg ladder under an OBC. It is easy to see that for a given $\{l_h\}$, one may still explicitly introduce the Marshall sign factor $ (-1)^{\hat{N} ^{\downarrow}_{A}} $ such that under its unitary transformation, the resulting $ \tilde{H}^{\beta}_J$ remains precisely the same ``sign-free'' form as in Eq. (\ref{tildeHJ}) with off-diagonal matrices always being negative-definite.

Let us first consider $H_0$ in Eq. (\ref{H0}), in which the hopping term Eq. (\ref{Hst}) can be transformed as follows:
\begin{align}\label{Hst1}
\tilde{ H}_{\sigma \cdot t_{\parallel}} \equiv  & (-1)^{\hat{N} ^{\downarrow}_{A}}  H_{\sigma \cdot t_{\parallel}}  (-1)^{\hat{N}^{\downarrow}_{A}}  \nonumber \\
= & - t\sum_{i\sigma} \left( c_{1i\sigma}^\dagger c_{1i+1\sigma}+ \ c_{2i\sigma}^\dagger c_{2i+1\sigma}+ \mathrm{h.c.} \right)
 \end{align}
whose off-diagonal matrices remain sign negative-definite in the Ising-spin-hole basis defined in Eq. (\ref{basis2}) under an OBC (such that no exchange between the doped holes occur, which would otherwise give rise to a fermion sign). Namely, the Perron-Frobenius theorem still applies to constructing a sign-free ground state for $H_0$ upon doping in the same form as in Eq. (\ref{phi0}).

The ground state governed by the sign-free ${H}_0$ defined in Eq. (\ref{H0}) thus looks quite conventional upon doping, which is studied by DMRG as given in the main text [cf. Fig. (2)]. There one finds that the two holes indeed form a bound state under $H_0$, once being injected into two distinct quantum spin states at $\beta\neq 0$. Here the pairing potential entirely comes from the original superexchange term $ \tilde{H}^{\beta}_J$ [cf. Eq. (\ref{tildeHJ})], which may be generally called the RVB-mechanism as proposed by Anderson~\cite{Anderson1987, Anderson}, in which the holes are paired up in order to minimize the energy cost of broken singlet/triplet spin bonds in $\tilde{H}^{\beta}_J $ caused by doping.

It is straight forward to show that the Cooper pair operators, defined in Eqs. (4) and (5) in the main text, can be transformed as follows:
\begin{align}\label{D}
\tilde{\Delta}_{ij}^{s,t} \equiv  & (-1)^{\hat{N} ^{\downarrow}_{A}} \hat{\Delta}_{ij}^{s,t} (-1)^{\hat{N}^{\downarrow}_{A}}  \nonumber \\
= & (-1)^{i-j}\hat{\Delta}_{ij}^{s,t}
 \end{align}
acting on the sign-free ground state $|{\Phi}_G\rangle_{0}$ in Eq. (\ref{phi0}). Here $(-1)^{i-j}$ is only a trivial sign factor depending on the definition of the Marshall sign rule in Fig.~\ref{sFig1} at either $\beta>0$ or $\beta<0$. Therefore, the \emph{sign-free} two-hole bound pair ground state of $H_0$ must be of the form
\begin{equation}\label{2h}
 |\Psi_G\rangle_{\emph{2h}}= \hat{\Delta} |\Psi_G\rangle_{\emph{half-filling}}
 \end{equation}
 where $\hat{\Delta}$ is a \emph{linear} combination of $\hat{\Delta}_{ij}^{s,t}$ with short-ranged $i-j$ at $\beta\neq 0$, where the half-filling spin background is always gapped. It means that the motion of such a hole pair must be always coherent or in other words, the pair-pair correlation should be long-ranged.

Two remarks are in order here. One is that by introducing the ``spin-orbit coupling'' in the hopping term Eq. (\ref{Hst}) of $H_0$, the spin rotational symmetry is no longer held in the two-hole state, such that the superscripts, $s$ and $t$, of the Cooper pair operator $\hat{\Delta}_{ij}^{s,t}$ no longer refer to the true singlet and triplet spin symmetries here. Second is that the pair-pair correlations of $\hat{\Delta}_{ij}^{s,t}$ as shown in Fig. 3 are calculated in a finite size with OBC such that a fall-down near the boundary is clearly shown there even though a long-range correlation is expected for Eq. (\ref{2h}) in an infinite sample.

\subsection{Emergent phase-string sign structure in $H$}

However, a nontrivial sign structure emerges in $H$ once holes are doped into the half-filling spin backgrounds. To see this, let us note that the hopping term  (\ref{Htj}) can be transformed as
\begin{align}\label{Ht2}
\tilde{ H}_{t_{\parallel}} \equiv  & (-1)^{\hat{N} ^{\downarrow}_{A}}  H_{t_{\parallel}}  (-1)^{\hat{N}^{\downarrow}_{A}}  \nonumber \\
= & - t\sum_{i\sigma}\sigma \left( c_{1i\sigma}^\dagger c_{1i+1\sigma}+ \ c_{2i\sigma}^\dagger c_{2i+1\sigma}+ \mathrm{h.c.} \right)
 \end{align}
whose off-diagonal matrices are generally not sign-definite due to the spin-dependent sign factor $\sigma=\pm 1$. It is precisely like $H_{\sigma \cdot t_{\parallel}}$ in Eq. (\ref{Hst}) in the original representation. Therefore, the Perron-Frobenius theorem no longer applies to constructing a sign-free ground state for $H$ upon doping. The aforementioned sign factor in Eq. (\ref{Ht2}) generally leads to the so-called phase string effect once the doped hole(s) starts to proliferate on the spin background~\cite{Sheng1996,Weng1997,Wu2008,Zaanen2011}, which results in strong non-perturbative effects including a novel pairing mechanism to be studied in this work.

In order to understand such a nontrivial sign structure, let us introduce the nonlocal unitary transformation
\begin{equation}\label{utransf}
e^{i\hat{\Theta}}\equiv e^{-i\sum_{\alpha i}n_{\alpha i}^h\hat{\Omega}_{\alpha i}}
\end{equation}
to formally remove the $\sigma$-sign appearing in $\tilde{H}_{t_{\parallel}} $  [Eq. (\ref{Ht2})]. Here,  $n_{\alpha i}^h$ denotes the hole number at leg $\alpha=1$, $2$ and site $i$ along the chain, and the nonlocal phase shift
\begin{equation}\label{Omega}
 \hat{\Omega}_{\alpha i}\equiv \pi \sum_{l>i}n_{\alpha l} ^{ \downarrow},
 \end{equation}
 which involves all $\downarrow$-spins (whose number operator is denoted by $n^{\downarrow}_{\alpha l}$) at sites $l>i$ under the OBC.

It is straightforward to check that
 \begin{equation}
 e^{-i\hat{\Theta}}  {H}_{t_{\parallel}}  e^{i\hat{\Theta}} \rightarrow { H}_{\sigma \cdot t_{\parallel}}
 \end{equation}
which becomes the sign free hopping term ${ H}_{\sigma \cdot t_{\parallel}}$ in $H_0$. Note that this is possible only for the case without the interchain hopping $t_{\perp}$ and under the OBC, which are invoked in the present study for simplicity.

On the other hand, ${H}^{\beta}_J $ [Eq. (\ref{eq:HJ})] is not invariant under the unitary transformation (\ref{utransf}), and it is easy to show
\begin{equation}
 e^{-i\hat{\Theta}}   {H}^{\beta}_J   e^{i\hat{\Theta}} \rightarrow {H}^{\beta}_J+{H}_I ~,
 \end{equation}
 where a new term ${H}_I$ is generated as follows:
\begin{align} \label{HI}
 {H}_I
 & = - \frac J 2 \beta \sum_{i} \left[1-e^{-i\pi \sum_{l<i}(n^h_{1l}-n^h_{2l})} \right] { S}^+_{1i}{S}^-_{2i}+h.c.
\end{align}
It means that the nontrivial sign structure introduced by doping cannot be truly ``gauged away'' by the unitary transformation. Instead, the irreparable ``phase string" effect shown in ${H}_{t_{\parallel}} $ of Eq. (\ref{Ht2}) is now precisely captured by ${H}_I$ in Eq. (\ref{HI}) above as an emergent nonlocal interaction.

Finally, the new Hamiltonian obtained under the unitary transformation is given by:
\begin{align}\label{H1}
H  \rightarrow  & ~ e^{-i\hat{\Theta}}  H e^{i\hat{\Theta}} \nonumber \\
& =  H_0+{H}_I ~.
 \end{align}
 Correspondingly, the ground state $|\Psi_G\rangle  $ is transformed into the new one $|\Phi_G\rangle$ as
 \begin{align}\label{Psi}
 |\Psi_G\rangle  =  e^{i\hat{\Theta}} |\Phi_G\rangle~.
 \end{align}
On the other hand, the nontrivial ``phase string'' properties of doping into the short-range ``spin liquid'' background are captured by ${H}_I$, whose singular role will be explored in the following.

\section{Emergent string-like confinement potential $H_I$ }

\subsection{Single hole case}

With the sign-free condition of $H_0$, one expects that the Bloch theorem still holds for excited doped holes governed by $H_0$ \emph{in the bulk} with the translational invariance, i.e., with the total momentum $k$ solely carried by the doped hole(s) moving in the spin gapped background. Of course, such an excitation does not contradictory to a ground state in which an edge mode with a doped hole trapped at one of the open boundaries, which can indeed occur in the single-hole doped case at $\beta<0$ [i.e., the AKLT regime, see Fig. ~\ref{Fig1}(b) in the main text].

Now we consider ${H}_I$ defined in Eq. (\ref{HI}). It is most straightforward to show that the string-like nature of ${H}_I$ leads to the breakdown of a hypothetic coherent quasiparticle (Bloch wave) motion of a single hole. Suppose a doped hole behaves like an extended Bloch wave with translation symmetry, then the mean value of $ \langle { S}^+_{1i}{S}^-_{2i}\rangle$ on each rung should not be changed, which approaches $ \langle { S}^+_{1}{S}^-_{2}\rangle_0$ at half-filling value at $L\rightarrow \infty$ (as a single extended hole has no thermodynamic effect). It then follows from Eq. (\ref{HI}) that the hole would experience a linear-confining potential $V$ in ${H}_I$,
\begin{align}\label{stringp}
V\equiv J\left |\beta  \langle { S}^+_{1}{S}^-_{2}\rangle_0+c.c. \right |\times  l_{i0},
\end{align}
where $l_{i0}$ is the length between the hole site $i$ and the right boundary on each chain of the two-leg ladder (noting that $\beta J \langle { S}^+_{1}{S}^-_{2}\rangle_0<0$). In other words, if one assumes that the single hole state is in a Bloch wave state, then ${H}_I$ should always lead to its instability towards a ``confinement'' as $\langle {H}_I\rangle \rightarrow V\propto J|\beta| L\rightarrow \infty$.

Here it is important to note that the form of the string-like potential in Eq. (\ref{stringp}), which is derived from Eq. (\ref{HI}) under the condition that the single hole is spatially uniformly distributed, is not necessarily rigid in general. Once the hole profile becomes localized or incoherent, the condition $ \langle { S}^+_{1i}{S}^-_{2i}\rangle \rightarrow \langle { S}^+_{1}{S}^-_{2}\rangle_0$ is no longer valid, and the whole self-trapping or incoherent solution must be determined self-consistently, which will be further discussed below.

 \subsection{Spontaneous translational symmetry breaking}

According to the above discussion, the instability of a \emph{coherent} extended hole state is inevitable due to the linear-potential in $H_I$. However, it does not specify a unique true ground state $ |{\Phi}_G\rangle_{1h}$ in which the hole is found localized only at the boundary. In the following, we will see that there can be multiple ground states, which may be explicitly revealed via nonlocal or large-gauge transformations.

To gain the insight of multi-solution nature of the ground state, one may introduce a \emph{different} nonlocal unitary transformation
to replace Eq. (\ref{Omega}) by
\begin{equation}\label{Omega1}
 \hat{\Omega}_{\alpha i}\equiv \pi \sum_{max \{i,i_0\}>l>min \{i,i_0\}}n_{\alpha l}^{ \downarrow} ~.
 \end{equation}
Then it is easy to show that the resulting ${H}_{0}+{H}_I $ will remain unchanged except that the ending point of the linear confining potential in Eq. (\ref{stringp}) is shifted from the right-hand-side boundary to an rung site $i_0$ along the chain direction. Consequently the hole would be confined near $i_0$ by $H_I$ instead of at the boundary in the ``ground state'' discussed in the previous subsection.

Physically, these multiple ``ground states'' imply the translational symmetry breaking. With the localization, the ground state $|\Psi_G\rangle_{1h}$ is degenerate with the confinement center $i_0$ located anywhere on the ladder at $L\rightarrow \infty$. It is important to point out that this is not contradictory to the translational invariance of the original Hamiltonian for the \emph{whole system} including all electrons. It just means that an injected hole propagating in the spin background has to exchange momentum with the latter, which is in the thermodynamic limit. As a result, a \emph{spontaneous} translational symmetry breaking occurs to the charge degree of freedom.  Note that for each choice of $i_0$, a \emph{bare} hole is supposed to be initially injected into the half-filled ground state at $i_0$ (with $e^{i\hat{\Theta}}=1$ at $n^h_{\alpha i_0}=1$).

It is also noted that the single hole trapped at an arbitray $i_0$, say, one of the boundary of the sample, is not necessarily the lowest energy state, or, the true ground state under an OBC, which explicitly breaks the translational symmetry. It so happens that at $\beta<0$ the edge modes have a slightly lower energy such that a DMRG calculation can easily pick up the ground state as a hole trapped at the boundary. Furthermore, a ground state may be also constructed by a superposition of those localized states uniformly distributed along the ladder's chain direction. Nevertheless, the general instability of a coherent Bloch wave solution $ |{\Phi}^0_G\rangle_{1h}$, once $H_I$ is included, indicates that the hole can no longer be described by a coherent excitation of a definite momentum.

\subsection{Enhanced binding energy due to  $H_I$ }

As already mentioned before, the DMRG calculations at $\beta=-1$ are shown in Fig. ~\ref{Fig1}(b) of the main text, which shows a single hole is well trapped at the open boundary of the two-leg ladder.
However, once two holes are injected into the spin ladder, as Fig. ~\ref{Fig1}(c) clearly shows, the deconfinement of the hole pair takes place. Indeed, according to Eq. (\ref{HI}), as the two holes sitting at the same rung $l$ of different chains, e.g., $n^h_{1l}=n^h_{2l}=1$, or at the nearest neighboring sites along the same chain, e.g., $n^h_{1l}=n^h_{1l+1}=1$, the confining potential disappears in ${H}_I$. That is, the same linear-potential force making a single hole localized/incoherent can be effectively removed if two holes are present in a tight pair. Or in other words, ${H}_I$ will provide a confining potential for two holes to be bound together to become a mobile object. It is easy to see that if the two holes are separated along the chain direction by a distance $l_{ij}$, then a pairing potential linearly proportional to $J|\beta|l_{ij}$, in a fashion similar to Eq. (\ref{stringp}), will arise.

Clearly, the predominant contribution of the hole pairing in $E_b$ mainly comes from that of $H_I$ or the phase string effect, while the RVB mechanism, i.e., $E_b^0$, is secondary as indicated in Fig. ~\ref{Fig2} of the main text. This pairing due to $H_I$ is non-BCS in nature: although the string-like pairing potential $H_I$ is originated from the superexchange term $H_J^{\beta}$ after a unitary transformation, it is really related to the sign structure in the kinetic energy term, i.e., ${ H}_{t_{\parallel}}$ in Eq. (\ref{Ht2}). In other words, the effect of $H_I$ would disappear at $t=0$ where ${ H}_{t_{\parallel}}$ is absent and ${H}^{\beta}_J $ provides the sole pairing force through the RVB mechanism. In this sense, the enhanced binding force of $H_I$ is by nature kinetic-energy-driven.


\begin{thebibliography}{99}
\bibitem{note1} We have also checked the possible singlet diagonal pair-pair correlation, which is much weaker compared with the singlet rung pair-pair correlation.
\bibitem{note2} With $t_{\perp}\neq 0$, the unitary transformation used in this work no longer directly applies to converting the $t$-$J$ model into a ``sign-free'' one. There a general phase string transformation will be needed.
\end{thebibliography}
\end{document}